\documentclass[12pt]{iopart}
\usepackage{latexsym}
\usepackage{graphics}
\usepackage{graphicx}
\usepackage{amssymb}
\usepackage[latin1]{inputenc}

\begin{document}

\title[AC susceptibility of polycrystalline LaFeAsO$_{0.94}$F$_{0.06}$]{Intergrain effects in the AC susceptibility of polycrystalline LaFeAsO$_{0.94}$F$_{0.06}$: comparison with cuprate superconductors}

\author{G. Bonsignore, A. Agliolo Gallitto and M. Li Vigni}
\address{CNISM and Dipartimento di Scienze Fisiche e Astronomiche, Università di Palermo, Via Archirafi 36, I-90123 Palermo (Italy)}

\author{J. L. Luo, G. F. Chen and N. L. Wang}
\address{Beijing National Laboratory for Condensed Matter Physics, Institute of Physics, Chinese Academy of Sciences, Beijing 100190, China}

\author{D. V. Shovkun}
\address{Institute of Solid State Physics, Russian Academy of Sciences, 142432 Chernogolovka, Moscow District}
\begin{abstract}
The AC susceptibility, $\chi$, at zero DC magnetic field of a polycrystalline sample of LaFeAsO$_{0.94}$F$_{0.06}$ ($T_c \approx 24$~K) has been investigated as a function of the temperature, the amplitude of the AC magnetic field (in the range $H_{ac} = 0.003 \div 4$~Oe) and the frequency (in the range $f = 10~\mathrm{kHz} \div 100~\mathrm{kHz}$). The $\chi(T)$ curve exhibits the typical two-step transition arising from the combined response of superconduncting grains and intergranular weak-coupled medium. The intergranular part of $\chi$ strongly depends on both the amplitude and the frequency of the AC driving field, from few Kelvin below $T_c$ down to $T=4.2$~K. Our results show that, in the investigated sample, the intergrain critical current is not determined by pinning of Josephson vortices but by Josephson critical current across neighboring grains.
\end{abstract}



\maketitle

\section{Introduction}
The recent discovery~\cite{kamihara} of superconductivity in iron-based oxypnictides, ReFeAsO$_{1-x}$F$_x$ with Re: rare earth, has attracted much attention by the condensed-matter-physics community. Such compounds exhibit magnetic order and superconductivity on varying the doping; furthermore, the critical temperature strongly depends on the fluorine content~\cite{Luo-Re}. In optimally doped samples, $T_c$ varies in the range 20 $\div$ 50~K depending on the Re ion. The main reason of this interest is the similarity between some properties of iron-based pnictides and those of cuprate superconductors, such as the layered structure, low carrier density, magnetic order in parent (undoped) compounds, possible unconventional pairing mechanism. Moreover, some studies have highlighted that the superconducting-gap structure of iron-based pnictides resembles that of MgB$_2$, with multi-gap superconductivity~\cite{kawasaki,samuely,Gonnelli-2gap,Ummarino}, which could originate from their multiple-electronic-band structures.

One of the characteristics of high-$T_c$ cuprates is related to their granularity~\cite{Clem-1988}; due to the small value of the coherence length, grain boundaries as well as defects even of small dimensions act as weak links, giving rise to flux penetration at low magnetic fields~\cite{kim-lam}, depression of the critical current density, nonlinear effects~\cite{Polich-fisicaC404,Noi_FisicaC161}. On the contrary, it has been extensively shown that in MgB$_2$ only a small number of grain boundaries act as weak links~\cite{Samanta}, giving rise to reduced nonlinear effects~\cite{Noi-Metamat}. In this scenario, it is of importance to explore granularity effects in iron-based oxypnictides and compare the results with those obtained in other superconductors.

Investigation of the electromagnetic response of different polycrystalline samples of iron-based oxypnictides have highlighted granularity effects, such as two-step susceptibility transition~\cite{polichetti}; different values of the global current density, flowing through weak-coupled grains, and the local current density, inside the superconducting grains~\cite{2scale_di_J}; small hysteresis in DC magnetization of bulk and powdered samples~\cite{granularityAPL}. However, it has not been clarified if such effects are due to intrinsic properties of the oxypnictides or are related to specific properties of the investigated samples.

Measurements of the AC susceptibility can be conveniently used to investigate both intergrain and intragrain vortex dynamics in superconductors~\cite{Clem-1988,gomory,libro-susc,Nikolo_Goldfarb,Muller_PhysC168,SUST_93,Polich-fisicaA,Chen_DX_APL2006}. In particular, measurements at low DC magnetic fields as a function of the AC-field amplitude allows to highlight  possible nonlinear effects related to critical state in the intergrain region; investigation as a function of the driving-field frequency allows to identify the different regimes of fluxon motion.

In this paper, we investigate the AC susceptibility of a sample of LaFeAsO$_{1-x}$F$_x$, with $x=0.06$, and compare the results with those obtained in ceramic cuprates. The measurements have been performed at zero DC magnetic field as a function of the temperature, for different values of the amplitude and the frequency of the driving field, and at $T=4.2$~K as function of the AC-field amplitude. The results are qualitatively discussed in the framework of the models reported in the literature for the vortex dynamics in the intergranular region. A qualitative comparison of the experimental $\chi(H_{ac}, f)$ curves with those obtained by Chen et \emph{al.}~\cite{Chen_DX_APL2006} in the framework of the so-called flux-flow critical-state model has highlighted that, in the investigated sample, the intergrain critical current is not determined by pinning of Josephson vortices but by Josephson critical current across neighboring grains.

\section{Experimental apparatus and sample}\label{sec:samples}
The AC susceptibility, $\chi = \chi^{\prime}+ i\chi^{\prime \prime}$,  has been measured in a polycrystalline sample of LaFeAsO$_{0.94}$F$_{0.06}$, prepared by solid state reaction using LaAs, Fe$_2$O$_3$, Fe and LaF$_3$ as starting materials. The sample has a nearly parallelepiped shape of approximate dimensions $1\times 2\times 2~\mathrm{mm}^3$; details on the preparation and properties of the sample are reported in Ref.~\cite{luo}. Atomic-Force-Microscopy (AFM) analysis on the sample surface has highlighted a high inhomogeneity; in particular, the sample seems to consist of grains of micrometric size embedded in a finer grained matrix.

The measurements have been performed by a computer controlled mutual-inductance susceptometer, which uses two identical secondary coils oppositely wounded. The coils are positioned coaxially inside a long primary coil, which provides an uniform AC field at the specimen position. The sample is located at the center of one of the secondary coils, by a sapphire holder, with the largest surface perpendicular to the coil axis. A temperature sensor and a heater are also fixed at the sapphire holder. The induced disbalance voltage in the secondary coils is proportional to the sample susceptibility: $v=\alpha f H_{ac} \chi$, where $\alpha$ depends on the sample and coil geometry. The signal is detected by a lock-in amplifier that allows to separate the real, $\chi^{\prime}$, and the imaginary, $\chi^{\prime \prime}$, components. To reduce the change of the background disbalance signal, the coils are maintained in a liquid He bath, while sample, sensor and heater are placed in gaseous He.

The AC susceptibility in the absence of DC magnetic field has been investigated as a function of the temperature, the amplitude of the AC driving field (in the range $H_{ac} \approx 0.003 \div 4$~Oe) and the frequency (in the range $f = 10~\mathrm{kHz} \div 100~ \mathrm{kHz}$). In order to disregard the constant $\alpha$, we have measured the $\chi^{\prime}$~vs~$H_{ac}$ curve at $T=4.2$~K; extrapolating the curve to $H_{ac}\rightarrow 0$, we have determined a constant, $\chi_0^{\prime}$, which has been used to normalize all the curves here reported.

\section{Experimental results}\label{experimental}
\begin{figure}[ht]
\centering
\includegraphics[width=8cm]{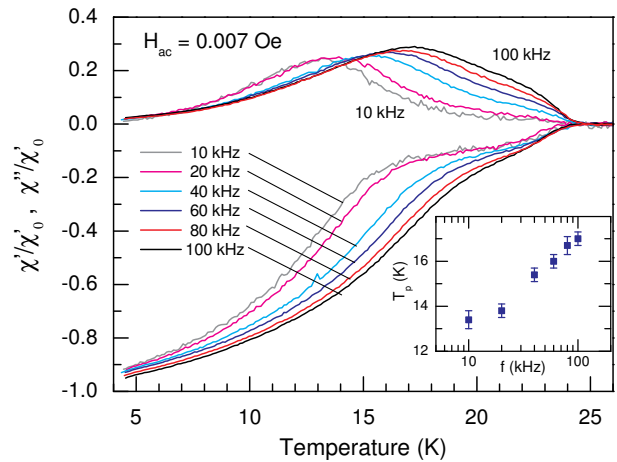}
\caption{Normalized values of $\chi^{\prime}$ and $\chi^{\prime \prime}$ as a function of the temperature, obtained at $H_{ac}=0.007$~Oe and different values of the frequency. Inset: dependence of $T_p$ on the driving-field frequency.}
\label{CHI(f)}
\end{figure}

Figure \ref{CHI(f)} shows the normalized value of $\chi^{\prime}$ and $\chi^{\prime \prime}$ as a function of the temperature, obtained at zero DC field and $H_{ac}=0.007$~Oe, at different values of the frequency (from 10~kHz to 100~kHz).
The $\chi^{\prime}(T)$ curves show the typical two-step transition due to the intergranular (at low $T$) and intragranular (near $T_c$) response~\cite{gomory,libro-susc}. In correspondence of the low-$T$ step, the $\chi^{\prime \prime}(T)$ curve shows a well-defined peak, at a characteristic temperature $T_p$, which should occur when the AC field penetrates through the intergranular region just to the center of the sample. On increasing the frequency, the peak intensity increases and $T_p$ moves towards higher temperatures. The kink at $T\sim 20$~K, well visible at high frequencies, should be related to the "intrinsic" $\chi^{\prime \prime}$ peak due to the intragrain field penetration. The inset shows the dependence of $T_p$ on the driving-field frequency.

The frequency dependence of $\chi$ reported in Figure~\ref{CHI(f)} is much more enhanced than that reported in the literature for high-$T_c$ cuprates~\cite{Nikolo_Goldfarb,Muller_PhysC168,SUST_93,Muller_PhysC197}. For example, Nikolo and Goldfarb~\cite{Nikolo_Goldfarb}, investigating the frequency dependence of AC susceptibility in ceramic YBCO, observed a shift of $T_p$ with the frequency, from 10 Hz to 1000 Hz, of few tens of K at small $H_{ac}$ and of the order of 1~K at $H_{ac}=10$~Oe. On the contrary, at $H_{ac}=0.007$~Oe, varying the frequency from 10~kHz to 100~kHz we observe a $T_p$ shift of about 4~K.

\begin{figure}[h]
\centering
\includegraphics[width=8cm]{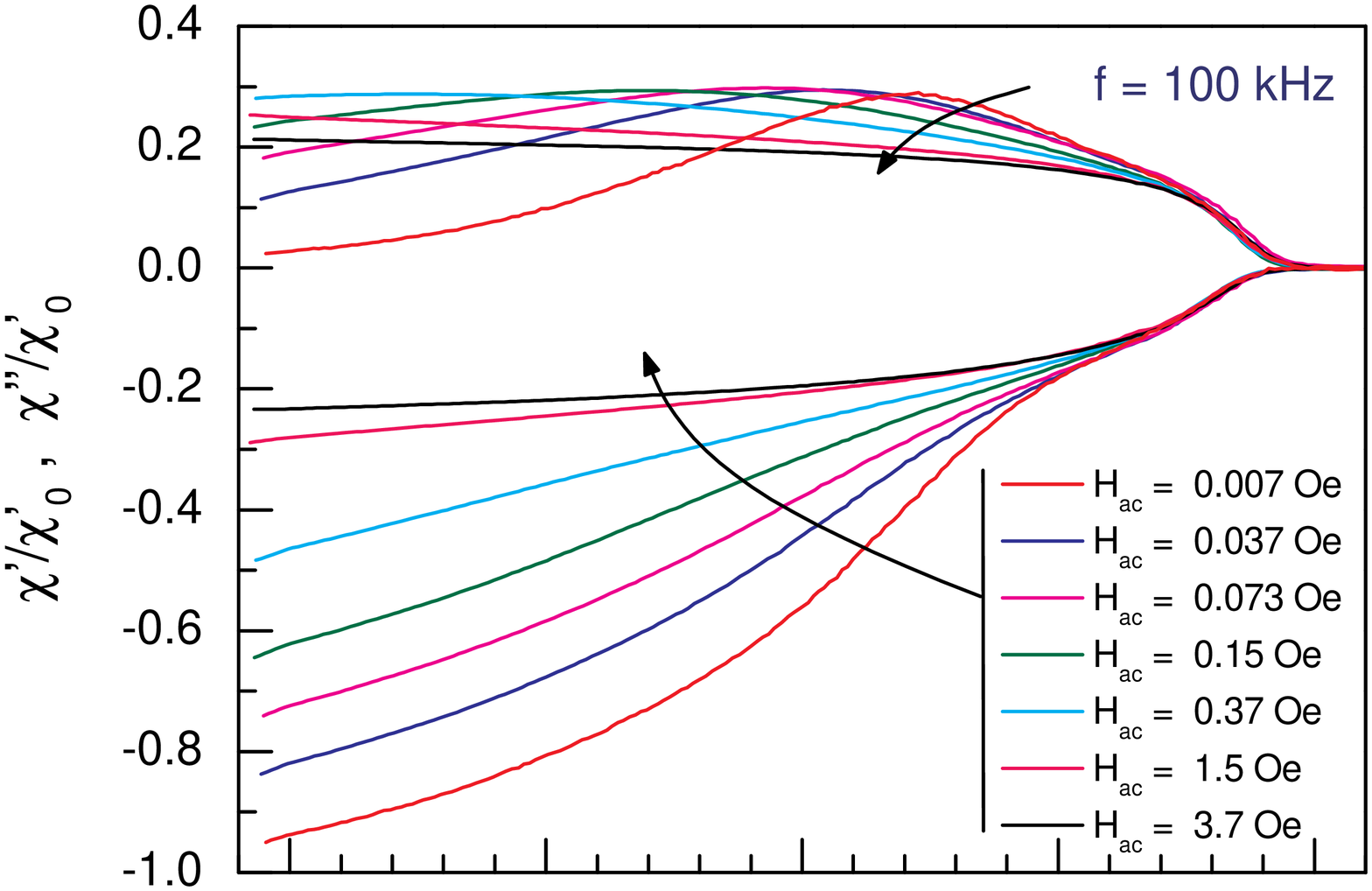}
\includegraphics[width=8cm]{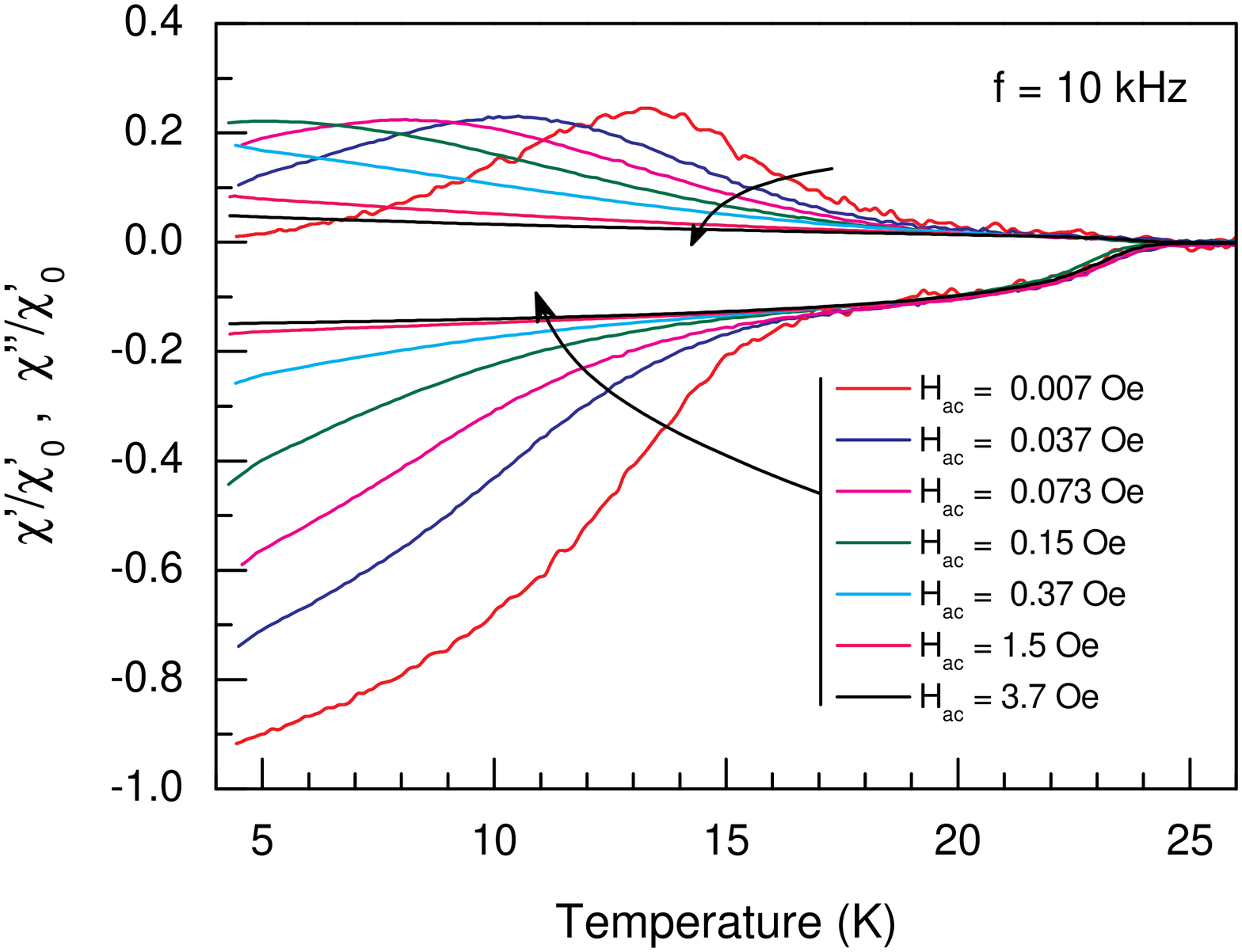}
\caption{Real and imaginary components of the AC susceptibility as a function of the temperature, obtained at zero DC field and different value of $H_{ac}$, for $f=100$~kHz and $f=10$~kHz. Arrows indicate increasing $H_{ac}$ values.}
\label{2Freq}
\end{figure}
Figure \ref{2Freq} shows the temperature dependence of $\chi^{\prime}$ and $\chi^{\prime \prime}$  at the two frequencies, $f=100$~kHz (a) and $f=10$~kHz (b), obtained at zero DC magnetic field and different values of the AC-field amplitude. One can note a near-$T_c$ region characterized by a linear response (independent of $H_{ac}$) and a region at lower $T$ highlighting enhanced nonlinear effects.

\begin{figure}[h]
\centering
\includegraphics[width=8cm]{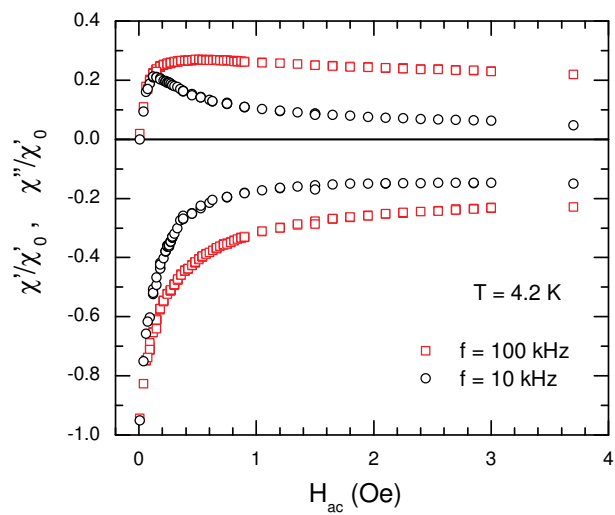}
\caption{$\chi^{\prime}$ and $\chi^{\prime \prime}$ as a function of the amplitude of the AC magnetic field, at two different values of the frequency.}
\label{CHI(Hac)}
\end{figure}

In Figure \ref{CHI(Hac)}, we report the dependence of $\chi^{\prime}$~(a) and $\chi^{\prime \prime}$~(b) on the amplitude of the AC field, obtained at $T=4.2$~K and at the two frequencies: $f=100$~kHz and $f=10$~kHz. Even at low temperatures, the susceptibility strongly depends on the frequency; in particular, the AC-field value at which $\chi^{\prime \prime}(H_{ac})$ exhibits the maximum increases of a factor of four on varying the frequency from 10~kHz to 100~kHz, and the maximum value of $\chi^{\prime \prime}/\chi_0^{\prime}$ varies from $\approx 0.21$ to $\approx 0.27$

\begin{figure}[h]
\centering
\includegraphics[width=7.8cm]{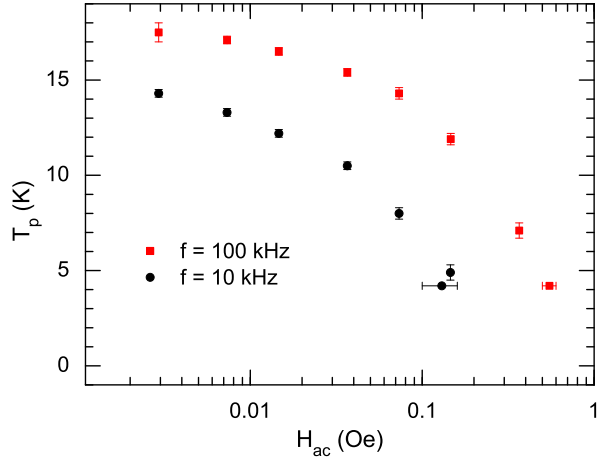}
\caption{Temperature position of the $\chi^{\prime \prime}$ peak as a function of the amplitude of the AC magnetic field, at two different values of the frequency. The points with horizontal error bars have been deduced from the data of Fig. 3}
\label{Tp(Hac)}
\end{figure}

By analyzing the results of Figures~\ref{2Freq} and \ref{CHI(Hac)} and other data not reported here, we have obtained the dependence of $T_p$ on the amplitude of the AC field; the results are shown in Figure~\ref{Tp(Hac)}. One can note a strong field dependence even at low $H_{ac}$ values; it is worth to remark that in high-$T_c$ cuprates $T_p$ nearly saturates for $H_{ac} \lesssim 0.1$~Oe~\cite{Muller_PhysC168,Muller_PhysC197}.

\section{Discussion}
Investigation carried out since the discovery of cuprate superconductors has shown that the AC susceptibility transition, with two steps in $\chi^{\prime}(T)$ accompanied by two peaks in $\chi^{\prime \prime}(T)$, is expected in samples consisting of superconducting grains interconnected by a system of weak links~\cite{gomory,libro-susc}. When cooled below $T_c$, the grains first become superconducting, and shielding is performed by the so-called intragrain current, with density $J_{cg}$. On further decreasing the temperature, the whole sample behaves as a homogeneous superconductor able to carry the macroscopic intergrain current density $J_{cm}$, which is determined by the Josephson current across neighboring grains. In this case, one expects to detect in $\chi^{\prime \prime}(T)$ a near-$T_c$ peak, which occurs when the AC field penetrates just to the center of grains, and a lower-$T$ peak corresponding to complete penetration of the AC field in the weak-link matrix. It is worth to remark that a near-$T_c$ peak, indicated as intrinsic peak would occur even in the case of reversible screening (negligible intragrain pinning); it is expected at the temperature at which the em-field penetration depth becomes comparable with the grain size.

From Figure~\ref{CHI(f)}, one can easily distinguish two steps in $\chi^{\prime}(T)$; according to the above considerations, the step near $T_c$ is ascribable to the superconducting-grain shielding, while that at lower $T$ to the intergrain shielding. In correspondence of the low-$T$ step, we observe a well defined peak in $\chi^{\prime \prime}(T)$, while the expected near-$T_c$ peak is not well resolved. We observe only a \emph{kink} at $T\approx 22$~K, better visible at high frequency; this finding could be due to a wide grain-size distribution and/or a wide distribution of $J_{cg}$. On the other hand, the inhomogeneity of grains does not hinder shielding inside each grain, whose effect is well visible in $\chi^{\prime}(T)$.

In order to understand if the near-$T_c$ response of grains is related to reversible or irreversible screening is convenient to look at the $\chi$ vs $T$ curves for different $H_{ac}$. From Figure~\ref{2Freq}, one can see that for $T\gtrsim 20$~K, the $\chi$ vs $T$ curves do not depend on $H_{ac}$ (linear response), while at lower $T$ enhanced nonlinear response is observed. Irreversible response during the AC-field cycle gives rise to nonlinear effects, which manifest themselves with components of $\chi$ oscillating at harmonic frequencies of the driving field as well as with dependence of $\chi^{\prime}$ and $\chi^{\prime \prime}$ on $H_{ac}$~\cite{Polich-fisicaC404,Noi_FisicaC161,libro-susc}. The results we obtained near $T_c$ strongly suggest that the intragrain response is related to reversible screening, because of weak intragrain pinning at $T\gtrsim 20$~K.

According to Chen et \emph{al.}~\cite{Chen_Cryo}, the measured sample susceptibility can be written as the sum of the intergranular susceptibility, $\chi_m$, and the grain susceptibility, $\chi_g$, weighted by the effective volume fractions of the intergranular region and superconducting grains (excluding a shell of thickness of the order of the field penetration depth).

At $T\ll T_c$ and for magnetic fields smaller than the lower critical field of grains it is expected that $\chi_g^{\prime}=-1$ and $\chi_g^{\prime \prime}=0$; so, the sample susceptibility results
\begin{equation*}
\chi^{\prime}=-f_g+(1-f_g)\chi_m^{\prime}\,,
\end{equation*}
\begin{equation*}
\chi^{\prime \prime}=(1-f_g)\chi_m^{\prime \prime}\,,
\end{equation*}
where $f_g$ is the volume fraction of the grains and $1-f_g$ that of the intergrain region.\\
Looking at Figures~\ref{2Freq} and~\ref{CHI(Hac)}, the variations of $\chi$ at low temperatures are ascribable to the intergrain contribution. On increasing $H_{ac}$, the field penetrates the intergrain region reducing the shielding until $\chi^{\prime}$ reaches a saturation value ascribable to the grain shielding. Correspondingly, $\chi^{\prime \prime}$ increases, reaches the maximum value when the field penetrates, through the intergrain medium, to the center of the sample, and then decreases. The results we obtained for $\chi^{\prime}$ allows to affirm that in the investigated sample the volume fraction of the superconducting grains able to shield is less than the 20\% of the whole sample volume; the presence of grains as small as the penetration depth reduces the shielding but contributes to the transport of the intergrain current.

\begin{figure}[b!]
\centering
\includegraphics[width=7.8cm]{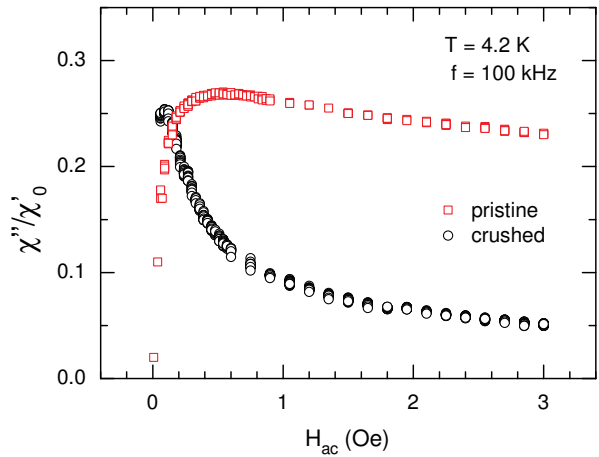}
\caption{Comparison between the $H_{ac}$ dependence of $\chi^{\prime \prime}$, obtained in the bulk and crushed samples.}
\label{confronto}
\end{figure}

Analysis of the results in the nonlinear regime can give useful information on the processes occurring in the intergrain region. In the framework of the Bean critical-state (CS) model, the value of $H_{ac}$ at which the $\chi^{\prime \prime}$ peak occurs roughly coincides with the full penetration field of the intergrain region, $H^*_m$. From Figure~\ref{CHI(Hac)} at $T=4.2$~K, we obtain $H_m^* \approx 0.55$~Oe at $f=100$~kHz and $H_m^* \approx 0.13$~Oe at $f=10$~kHz. These values are very small and should indicate extremely low values of the intergrain critical current density or intergrain-current path around regions much smaller than the sample dimension. In order to check if the latter hypothesis applies, we have coarsely crushed a portion of the sample, obtaining a powder with fragments of dimensions ranging from 50~$\mu$m to 500~$\mu$m. Figure~\ref{confronto} shows $\chi^{\prime \prime}$ as a function of the AC-field amplitude in the pristine and crushed samples, at $T=4.2$~K and $f=100$~kHz. One can see that $H^*_m$ in the crushed sample is about one order of magnitude smaller than that obtained in the pristine sample. Since the ratio of the two $H^*_m$ is of the same order of that between the dimension of the bulk sample and the mean size of the powder grains, we can rule out the possibility to have intergrain-current path much smaller than the whole bulk sample (i.e. the weak-link matrix extents over the whole sample).

It is worth noting that we observe a strong frequency dependence of both the height and the position of the $\chi^{\prime \prime}$ peak, which is not expected in the mere CS. Frequency dependence of the AC susceptibility has been discussed by different authors in the framework of different models~\cite{Muller_PhysC168,Polich-fisicaA,Chen_DX_APL2005,Rao,Ozogul}. Muller~\cite{Muller_PhysC168} considers the granular superconductor as a regular array of Josephson junctions connecting cubic grains, and Josephson vortices are assumed to hop between adjacent pinning centers located at the grain corners. In the framework of this model, the frequency dependence is ascribed to effects of magnetic relaxation during the AC cycle and it is taken into account by adding a flux-creep term to the current term in the CS equation. Other authors~\cite{Rao,Ozogul} take into account the magnetic relaxation by assuming a proper frequency dependence of the intergranular critical current.

Polichetti et \emph{al.}~\cite{Polich-fisicaA} have solved the nonlinear equation describing the diffusion of the magnetic field inside the sample, hypothesizing a proper dependence of the pinning potential on the density current, $U(J)$. Numerical computation of AC susceptibility using different $U(J)$ laws, corresponding to the linear dependence in the Kim-Anderson model and the nonlinear ones in the vortex-glass models, has highlighted a way to distinguish the different vortex phases. In particular, on increasing the frequency the height of the loss peak in the $\chi^{\prime \prime}(T)$ curve is expected to increase in the vortex-glass phase and to decrease in the Kim-Anderson case~\cite{Polich-fisicaA}. From Figure~\ref{CHI(f)}, one can see that on increasing the frequency the height of the $\chi^{\prime \prime}(T)$ peak increases, suggesting that the intergrain-vortex lattice could be in the vortex-glass phase.

Although the above-mentioned models predict a $\chi^{\prime \prime}(T)$-peak shift towards higher temperatures with the frequency, the shift we observe is much more enhanced than the expected one, no matter the creep exponent used. Moreover, we observe a strong dependence of the susceptibility on both the amplitude and frequency of the AC field even at low temperature (see Fig.~\ref{CHI(Hac)}) where thermal activated flux-creep should be slightly effective. So, our results cannot fully be explained by considering flux-creep effects.

Chen et \emph{al.}~\cite{Chen_DX_APL2006} have investigated the AC susceptibility of a sintered YBCO superconductor as a function of the amplitude and frequency of the driving field at fixed temperatures. In order to recognize the mechanisms determining the intergranular critical current at the different frequencies, they have numerically calculated $\chi(H_{ac},f)$ using different $E(J)$ characteristics, in the different motion regimes (collective flux creep, Kim-Anderson flux creep and flux flow). By using, for the flux-flow case, $E(J)=[J-sgn(J)J_c]\rho_f$ occurring at $|J|>J_c$ with $\rho_f$ the flux-flow resistivity, the authors show that the experimental results are consistent with those obtained by the so-called \emph{flux-flow critical-state} (FFCS) model~\cite{Chen_DX_APL2005}. In this model, $J_{cm}$ arises directly from the Josephson currents across weak-linked grains and the flux-flow behavior should be a consequence of the resistively shunted Josephson-junction network.

\begin{figure}[h]
\centering
\includegraphics[width=8cm]{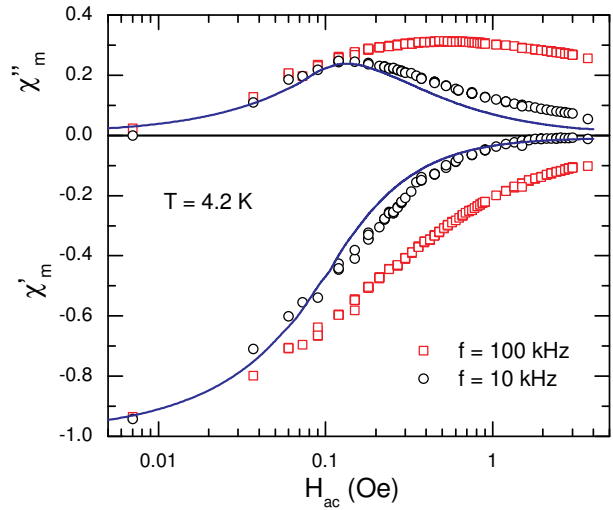}
\caption{Symbols: real and imaginary component of the integranular AC susceptibility of the LaFeAsO$_{0.94}$F$_{0.06}$ sample, $\chi_m^{\prime}$ and $\chi_m^{\prime \prime}$, as a function of the amplitude of the AC magnetic field, obtained at $T=4.2$~K for $f=10$~kHz and $f=100$~kHz; the results have been obtained from the data of Fig.~\ref{CHI(Hac)} supposing the volume fraction of grains to be $f_g=0.14$. Continuous lines show the results expected in the Bean CS model with $J_c=1\times 10^4$~A/m$^2$.}
\label{CHI(logHac)}
\end{figure}

The $\chi(H_{ac}, f)$ curves obtained by the FFCS model  have different features with respect to those obtained from the flux-creep models. In particular: i) the region of the curves for $H_{ac}$ values smaller than the value at which $\chi^{\prime \prime}$ exhibits the maximum are slightly affected by the frequency variations; ii) on increasing the frequency, the maximum of the $\chi^{\prime \prime}(H_{ac})$ curve shifts towards higher $H_{ac}$, its value increases and the peak broadens; iii) on increasing the frequency, $\chi^{\prime}(H_{ac})$ grows more slowly with $H_{ac}$, reaching its saturation value at higher $H_{ac}$. In Figure~\ref{CHI(logHac)}, we report as symbols the susceptibility of the LaFeAsO$_{0.94}$F$_{0.06}$ sample as a function of $\log(H_{ac})$, properly normalized to extract $\chi_m$; in particular, we have set the volume fraction of the grains $f_g=0.14$. Comparing these curves with those reported in Figure~2 of Ref.~\cite{Chen_DX_APL2006}, one can note that $\chi_m(H_{ac}, f)$ curves we obtained are consistent with the results expected from the FFCS model.

An important result highlighted by Chen et \emph{al.} is that the intergrain susceptibility calculated by the FFCS model in the low-frequency limit coincides with that obtained by using the Bean CS model. Since the experimental $\chi^{\prime \prime}(H_{ac})$-peak value at 10~kHz is compatible with that expected in the Bean CS model, we have calculated the $\chi_m(H_{ac})$ curves by using the Bean CS model; to take into account the demagnetization effects, we have set $H_{ac}^{eff}= H_{ac}-N M_m-n M_g$, where N (n) is the demagnetization factor related to the shape of the sample (grains) and $M_m$ ($M_g$) is the intergrain (grain) magnetization. In order to consider the grain demagnetization effects, the grains have been supposed to have spherical shape ($n=1/3$) and to completely screen the field ($M_g=-f_g H_{ac}^{eff}$). The contribution to $H_{ac}^{eff}$ of the intergrain matrix has been taken into account by a procedure that iteratively calculates $H_{ac}^{eff}$, considering the volume susceptibility of the matrix, until two consecutive values differ less than 1\%. The $\chi_m(H_{ac})$ curves calculated by this procedure are shown in Figure~\ref{CHI(logHac)} as continuous lines; they have be obtained with $J_c=1\times 10^4$~A/m$^2$ independent of $H_{ac}$. As one can see, the calculated $\chi_m(H_{ac})$ curves account quite well for the experimental ones up to  $H_{ac}\approx$~0.2 Oe. The small discrepancy at higher AC fields between the experimental 10~kHz curves and the calculated ones may be due to fact that the low-frequency limit is not attained in the sample. We have also calculated $\chi_m$ supposing an exponential field dependence of $J_c$, but it gave a worse fit. The 100 kHz curves are consistent with those obtained by Chen et \emph{al.}~\cite{Chen_DX_APL2005,Chen_DX_APL2006} by the FFCS model in the eddy-current limit, with a broad and more intense $\chi_m^{\prime \prime}(H_{ac})$ peak and lower values of $|\chi^{\prime}(H_{ac})|$ with respect to the 10~kHz curves.

The consistency of our experimental results with the theoretical ones obtained by Chen et \emph{al.} in the framework of the FFSC model suggests that in the investigated sample the intergrain critical current is not determined by pinning of Josephson vortices but by the Josephson critical current across grain boundaries. It would be interesting to understand if this is related to specific properties of the investigated sample or to intrinsic properties of iron-based oxypnictides. To this aim, samples of LaFeAsO$_{1-\delta}$F$_{\delta}$ with different fluorine content, and/or prepared by different methods, should be studied by techniques able to investigate the intergrain properties.

\section*{Conclusion}
We have investigated the AC susceptibility of a sample of LaFeAsO$_{0.94}$F$_{0.06}$, in the absence of static magnetic field, in the range of frequency $f=$10 kHz $\div$ 100 kHz. The real and imaginary components of $\chi$ have been measured as a function of the temperature, at different values of the amplitude and the frequency of the AC field, and at $T=4.2$~K as function of the AC-field amplitude, at the two frequencies $f=$~10~kHz and $f=$~100~kHz.

The $\chi(T)$ curves exhibit a two-step transition due to the intergranular and intragranular response. Measurements at different $H_{ac}$ have highlighted a linear response at $T\gtrsim 20$~K, ascribable to reversible screening of the superconducting grains. The results obtained for lower temperatures have highlighted that the intergrain susceptibility strongly depends on the amplitude and frequency of the AC field. On increasing $H_{ac}$ the intergrain shielding reduces and approaches its saturation value, corresponding to the complete shielding of grains, at applied field of the order of few Oe. This allowed us to estimate the volume fraction of the superconducting grains able to shield the field to be roughly  of 15\% of the whole sample volume. This finding can be ascribed to the presence in the sample of very small grains, as highlighted by AFM measurements.

Analysis of the results at low temperatures have been qualitatively discussed in the framework of models reported in the literature for the electrodynamic response of the intergrain medium. The features of the intergranular $\chi(H_{ac},f)$ curves are consistent with the results obtained in the framework of the so-called flux-flow critical-state model, developed by Chen et \emph{al.} for describing the electromagnetic response of the intergrain medium. In particular, we have shown that the $\chi(H_{ac})$ curves obtained at 10~kHz are in qualitative agreement with the results expected from the Bean critical-state model, consistently with the results obtained by Chen et \emph{al.} in the low-frequency limit; the experimental results obtained at 100~kHz are consistent with those obtained, by the same model, in the flux-flow regime. This finding strongly suggests that in the investigated LaFeAsO$_{0.94}$F$_{0.06}$ sample the intergranular critical current is determined by the maximum Josphson current across neighboring grains but not by pinning of Josepson vortices.

\ack
The authors are very glad to thank Massimiliano Polichetti and Marco Bonura for the interest on this work and fruitful suggestions. Work partially supported by the University of Palermo in the framework of the International Co-operation Project CoRI 2007 Cupane.


\section*{References}
\begin{thebibliography}{99}
\bibitem{kamihara} Kamihara Y, Watanabe T, Hirano M and Hosono H 2008 \emph{J. Am. Chem. Soc.} \textbf{130} 3296

\bibitem{Luo-Re} Chen G F, Li Z, Wu D, Dong J, Li G, Hu W Z, Zheng P, Luo J L and Wang N L 2008 \emph{Phys. Rev. Lett.} \textbf{101} 57007

\bibitem{kawasaki}Kawasaki S, Shimada K, Chen G F, Luo J L, Wang N L and Guo-qing Zheng 2008 \emph{Phys. Rev} B \textbf{78} 220506 (R)

\bibitem{samuely} Samuely P, Szab$\acute{\mathrm{o}}$ P, Pribulov$\acute{\mathrm{a}}$ Z, Tillman M E, Bud'ko S and Canfield P C 2009 \emph{Supercond. Sci. Technol.} \textbf{22} 014003

\bibitem{Gonnelli-2gap}Gonnelli R S, Daghero D, Tortello M, Ummarino G A, Stepanov V A, Kim JS and Kremer R K 2009 \emph{Phys. Rev.} B \textbf{79} 184526

\bibitem{Ummarino}Ummarino G A, Tortello M, Daghero D and Gonnelli R S 2009 \emph{Phys. Rev.} B \textbf{80} 172503

\bibitem{Clem-1988}Clem J R 1988 \emph{Physica} C \textbf{153-155} 50

\bibitem{kim-lam}Kim Y. Lam G H and Jeffries C D 1991 \emph{Phys. Rev.} B \textbf{43} 11404

\bibitem{Polich-fisicaC404} Adesso M G, Senatore C, Polichetti M and Pace S 2004 \emph{Physica} C \textbf{404} 289

\bibitem{Noi_FisicaC161} Ciccarello I, Guccione M and Li Vigni M 1989 \emph{Physica} C \textbf{161} 39

\bibitem{Samanta}Samanta S B, Narayan H, Gupta A, Narlikar A V, Muranaka T and Akimitsu J 2002 \emph{Phys. Rev.} B. \textbf{65} 092510

\bibitem{Noi-Metamat}Agliolo Gallitto A, Bonsignore G, Di Gennaro E, Giunchi G, Li Vigni M, and Manfrinetti P 2006 \emph{Microwave Opt. Technol. Lett.} \textbf{48} 2482

\bibitem{polichetti} Polichetti M, Adesso M G, Zola D, Luo J L, Chen G F, Li Z, Wang N L, Noce C, and Pace S 2008 \emph{Phys. Rev.} B \textbf{78} 224523

\bibitem{2scale_di_J}Yamamoto A, Polyanskii A A, Jiang J, Kametani F, Tarantini C, Hunte F, Jaroszynski J, Hellstrom E E, Lee P J, Gurevich A, Larbalestier D C, Ren Z A, Yang J, Dong X L, Lu W and Zhao Z X 2008 \emph{Supercond. Sci. Technol.} \textbf{21} 095008

\bibitem{granularityAPL} Yamamoto A, Jiang J, Tarantini C, Craig N, Polyanskii A A, Kametani F, Hunte F, Jaroszynski J, Hellstrom E E, Larbalestier D C, Jin R, Sefat A S, McGuire M A, Sales B C, Christen D K and Mandrus D 2008 \emph{Appl. Phys. Lett.} \textbf{92} 252501

\bibitem{gomory}Gömöry F 1997 \emph{Supercond. Sci. Technol.} \textbf{10} 523

\bibitem{libro-susc} Hein R A, Francavilla T L and Liebenberg D H 1991 \emph{Magnetic susceptibility of superconductors and other spin systems} (New Work: Plenum)

\bibitem{Nikolo_Goldfarb} Nikolo M and Goldfarb R B 1989 \emph{Phys. Rev.} B \textbf{39} 6615

\bibitem{Muller_PhysC168} Muller K H 1990 \emph{Physica C} \textbf{168} 585

\bibitem{SUST_93}Dhingra I and Das B K, 1993 \emph{Supercond. Sci. Technol.} \textbf{6} 765

\bibitem{Polich-fisicaA}Polichetti M, Adesso M and Pace S 2004 \emph{Physica} A \textbf{339} 119

\bibitem{Chen_DX_APL2005}Chen D -X, Pardo E and Sanchez A 2005, \emph{Appl. Phys. Lett.} \textbf{86} 242503

\bibitem{luo} Chen G F, Li Z, Li G, Zhou J, Wu D, Dong J, Hu W Z, Zheng P, Chen Z J, Luo J L and Wang N L 2008 \emph{Phys. Rev. Lett.} \textbf{101} 57007

\bibitem{Muller_PhysC197} Savvides N, Katsaros A, Andrikidis C and Muller K H 1992 \emph{Physica} C \textbf{197} 267

\bibitem{Chen_Cryo} Chen D -X, Nogues J and Rao K V 1989 \emph{Cryogenics} \textbf{29} 800

\bibitem{Rao} Jönsson B J, Rao K V, Yun S H and Karlsson U O 1998 \emph{Phys. Rev.} B \textbf{58} 5862

\bibitem{Ozogul}Ozogul O 2005 \emph{Phys. Stat. Sol.} \textbf{202} 1793

\bibitem{Chen_DX_APL2006}Chen D -X, Pardo E, Sanchez A and Bartolomé E 2006, \emph{Appl. Phys. Lett.} \textbf{89} 072501

\end{thebibliography}
\end{document}